\documentclass[journal]{IEEEtran}
\ifCLASSINFOpdf
  \usepackage[pdftex]{graphicx}
  \graphicspath{ {figs} }
  \DeclareGraphicsExtensions{.pdf,.jpeg,.png}
\else
\fi
\usepackage[table,xcdraw]{xcolor}
\usepackage{tabularx}
\usepackage{multirow}
\usepackage{amsmath} % added by lhh

\usepackage[]{changes}
\definechangesauthor[name={liangxiu han},color= blue]{LHAN}
\usepackage{csquotes}
\usepackage{todonotes}
\usepackage[switch]{lineno}

\begin{document}
\title{CXR-Net: An Encoder-Decoder-Encoder Multitask Deep Neural Network for Explainable and Accurate Diagnosis of COVID-19 pneumonia with Chest X-ray Images}


\author{Xin Zhang, Liangxiu Han*, Tam Sobeih, Lianghao Han, Nina Dempsey,  Symeon Lechareas, Ascanio Tridente, Haoming Chen, Stephen White

\thanks{ Xin Zhang, Liangxiu Han, Tam Sobeih: The Department of Computing, and Mathematics, Manchester Metropolitan University, Manchester M15GD, U.K (e-mail: x.zhang@mmu.ac.uk; l.han@mmu.ac.uk;T.Sobeih@mmu.ac.uk)}
\thanks{ Lianghao Han: The Department of Computer Science,  Brunel University, Uxbridge UB8 3PH, U.K (e-mail:lianghao.han@brunel.ac.uk)}

\thanks{ Nina Dempsey: The Department of Life Sciences, Manchester Metropolitan University, Manchester M15GD, U.K (e-mail:n.dempsey-hibbert@mmu.ac.uk) }
\thanks {Symeon Lechareas: The Radiology, Whiston Hospital, St Helens and Knowsley Teaching Hospitals NHS Trust, Warrington Road, Prescot L355DR, U.K}
\thanks{ Ascanio Tridente: The Intensive Care Unit, Whiston Hospital, St Helens and Knowsley Teaching Hospitals NHS Trust, Warrington Road, Prescot L355DR, U.K (e-mail:Ascanio.Tridente@sthk.nhs.uk)}
\thanks{ Haoming Chen: The Computer Science and Artificial
Intelligence,University of Sheffield, Sheffield S14DP, U.K}
\thanks{ Stephen White: The Cardiovascular Pathology,
Manchester Metropolitan University, Manchester M15GD, U.K (e-mail:stephen.white@mmu.ac.uk)}

\thanks {Corresponding author*: L. Han (e-mail: l.han@mmu.ac.uk)}

}

%\author{Xin Zhang, Liangxiu Han, Daoqiang Zhang
       
%\thanks{xxx was with the Department
%of Electrical and Computer Engineering, Georgia Institute of Technology, Atlanta,
%GA, 30332 USA e-mail: (see http://www.michaelshell.org/contact.html).}}% <-this % stops a space


\markboth{Journal of \LaTeX\ Class Files,~Vol.~14, No.~8, August~2015}%
{Shell \MakeLowercase{\textit{et al.}}: Bare Demo of IEEEtran.cls for IEEE Journals}



\maketitle

\begin{abstract}
Accurate and rapid detection of COVID-19 pneumonia is crucial for optimal patient treatment. Chest X-Ray (CXR) is the first line imaging test for COVID-19 pneumonia diagnosis as it is fast, cheap and easily accessible. Inspired by the success of deep learning (DL) in computer vision, many DL-models have been proposed to detect COVID-19 pneumonia using CXR images. Unfortunately, these deep classifiers lack the transparency in interpreting findings, which may limit their applications in clinical practice. The existing commonly used visual explanation methods are either too noisy or imprecise, with low resolution, and hence are unsuitable for diagnostic purposes. In this work, we propose a novel explainable deep learning framework (CXRNet) for accurate COVID-19 pneumonia detection with an enhanced pixel-level visual explanation from CXR images. The proposed framework is based on a new Encoder-Decoder-Encoder multitask architecture, allowing for both disease classification and visual explanation. The method has been evaluated on real world CXR datasets from both public and private data sources, including: healthy, bacterial pneumonia, viral pneumonia and COVID-19 pneumonia cases The experimental results demonstrate that the proposed method can achieve a satisfactory level of accuracy and provide fine-resolution classification activation maps for visual explanation in lung disease detection. The Average Accuracy, the Precision, Recall and F1-score of COVID-19 pneumonia reached 0.879, 0.985, 0.992 and 0.989, respectively. We have also found that using  lung segmented (CXR) images can help improve the performance of the model. The proposed method can provide more detailed high resolution visual explanation for the classification decision, compared to current state-of-the-art visual explanation methods and has a great potential to be used in clinical practice for COVID-19 pneumonia diagnosis. 

\end{abstract}


% \fi
\IEEEpeerreviewmaketitle

\section{Introduction}
\IEEEPARstart{S}ince December 2019, the world is experiencing a global pandemic from the emergence and spread of the potentially fatal COVID-19 (COronaVIrus Disease 2019) caused by severe acute respiratory syndrome coronavirus 2 (SARS-CoV-2) \cite{Taylor2020Review}. To optimize the management of this disease, accurate and rapid diagnosis is essential.

\par Medical imaging is used extensively in the diagnosis of COVID-19 CXR and Computer tomography (CT) are the main imaging modalities used to diagnose the respiratory complications of the virus\cite{Hassanien2020Automatic,Kadry2020Development,Sethy2020Detection,Holshue2020First}. Typical imaging features of COVID-19 pneumonia on CT include extensive consolidation, ground-glass opacity (GGO), typical of acute lung injury, lung consolidation, bilateral patchy shadowing, pulmonary fibrosis, multiple lesions and crazy-paving pattern etc,\cite{Holshue2020First,Orioli2020COVID-19}. CXR can also be used for COVID-19 pneumonia detection. Compared to CT, CXR has some advantages; it is faster, cheaper and does not expose the patient to the same levels of radiation \cite{Sethy2020Detection}. However, CXR has been shown to have poor sensitivity and positive predictive value for the detection of COVID-19 related pulmonary opacity, when compared to chest CT. Most people with COVID-19 infection do not develop pneumonia. However, pneumonia is much more frequent in those patients who become critically ill with COVID-19. A rapid, cheap and easily interpretable imaging test would be extremely useful for this subset of patients and so improving the sensitivity of such a technique would be extremely beneficial.

%\par Besides the laboratory-based tests, such as nucleic acid, antigens and serology tests etc., medical imaging is another major approach for detecting the COVID-19 \cite{Hassanien2020Automatic,Kadry2020Development,Sethy2020Detection,Holshue2020First}. It can include computed tomography (CT) scans \cite{Hassanien2020Automatic}, Chest X-ray (CXR), Camera Technology, Ultrasound Technology(US) and Radio Frequency (RF) signal sensing \cite{Kadry2020Development}. Among them, chest computed tomography (CT) is often used as a complementary examination in the diagnosis and management of COVID-19. Typical imaging features of COVID-19 on CT include extensive consolidation, ground-glass opacity (GGO), typical of acute lung injury, lung consolidation, bilateral patchy shadowing, pulmonary fibrosis, multiple lesions and crazy-paving pattern etc.\cite{Holshue2020First,Orioli2020COVID-19}. CXR can also be used for COVID-19 detection. Compared to CT, CXR has some advantages, such as faster, cheaper, and low risk from radiation hazards to human health \cite{Sethy2020Detection}. However, CXR demonstrated poor sensitivity and positive predictive value in detecting pulmonary opacities, a morphological pattern of Covid-19 on chest CT images. Using CXR to identify pneumonia may lead to significant rates of misdiagnosis \cite{Orioli2020COVID-19}. Most people with COVID-19 infection do not develop pneumonia. However, CXR of seriously ill patients with respiratory symptoms can help to identify the existence of COVID-19 pneumonia. 

\par Over the past several years, deep learning (DL) methods have shown enormous potential in medical imaging analyses for disease detection. With more publicly available CXR images, various deep learning-based methods have been extensively used to assist in the diagnosis of lung related diseases including Covid-19 \cite{Basu2020Deep,Feng2019Deep}. These methods are broadly divided into three categories: Lung disease classification \cite{Cheng2020Panoptic-DeepLab:,Hasan2020CVR-Net:,Shelke2020Chest}, deep learning-based explanation\cite{Ahsan2020Study,Majkowska2019Chest,Tang2020Automated,Wang2020COVID-Net:} and Organ area segmentation\cite{Fan2020Inf-Net:,Gaal2020Attention,Saeedizadeh2020COVID,Selvan2020Lung}. For the lung disease classification, the most commonly used deep learning based classifiers, such as AlexNet, DenseNet, ResNet etc., have been used for the lung disease classification on CXR images and have achieved reasonably high accuracy \cite{Rajpurkar2017CheXNet:,Rajpurkar2018Deep}. However, due to the black box nature in deep learning models, they do not provide the explanation to the classification results, leading to a limited understanding of the predictions. Some researchers have tried to use existing artificial intelligence explanation methods to interpret the classification of results on CXR images \cite{Ahsan2020Study,Brunese2020Explainable,Ghoshal2020Estimating,Zhang2020Viral}. However, the results from these methods were either too noisy or imprecise with low resolution, and hence not suitable for medical diagnostic purposes. To improve the accuracy of the model, some researchers also used deep learning-based segmentation methods for organs and lesions to reduce the effect of disease-unrelated image information on CXR images \cite{Gaal2020Attention,Saeedizadeh2020COVID}, but with little effort on improving interpretation.
To address the aforementioned limitations, particularly with respect to understanding the classification of results, the current work presents an explainable classification model for COVID-19 pneumonia characterization with enhanced visualization of feature representations, named as CXRNet. The contributions of this paper include:

\begin{enumerate}
  \item A novel deep learning based COVID-19 pneumonia diagnosis framework has been designed and implemented, allowing for both accurate disease classification and pixel-level visual explanation. This framework consists of two main components: a CXR image pre-processing module with lung segmentation and image enhancement, and a multitask deep learning model, named as CXRNet.
  \item A novel Encoder-Decoder-Encoder based multitask architecture has been designed in the proposed model. It consists of two jointly trained encoders for classification. The first one is used to learn the features of original CXR images and the second one is used to learn the features of the representation image generated from the decoder. The representation image between the two encoders acts as a proxy to visualise the most relevant infected areas for COVID-19 pneumonia diagnosis.
  \item An image preprocessing approach with Lung segmentation and Image enhancement units is used to improve the classification accuracy of the proposed model.
\end{enumerate}
\par The remainder of this paper is organized as follows: Section II reviews the related work. Section III describes the Datasets and methods; Section IV presents the experiment design and results; and Section V concludes the work.

\section{Related work}
\par In this section, we review three types of deep learning-based applications related to our work, including Lung disease classification, Lung segmentation and Artificial intelligence explanation.
\subsection{Lung Disease Classification}
\par The deep learning-based lung disease classification is the most commonly used application on CXR images. Pranav Rajpurkar et al. \cite{Rajpurkar2018Deep} developed a 121-layer DenseNet model \cite{Huang2018Densely}, called CheXNeXt, to detect the presence of 14 different thoracic pathologies using frontal view chest radiographs, such as pneumonia, pleural effusion, pulmonary masses and nodules etc. The model was trained and validated on the "ChestX-ray8" dataset \cite{Wang2017ChestX-ray8:}, which was a public repository of Chest X-ray images including 108,948 frontal view X-Ray images with eight lung disease labels (Atelectasis, Cardiomegaly, Effusion, Infiltration, Mass, Nodule, Pneumonia and Pneumothorax). Ivo M. Baltruschat et al. \cite{Baltruschat2019Comparison} used the same datasets to investigate the depth sensitivity of deep learning models on lung disease classification with CXR images by testing Resnet \cite{He2015Deep} with three different depths (layers), 34, 50 and 101.
The deep learning-based classifiers have also been used for the diagnosis of COVID-19. Narin et al. \cite{Narin2020Automatic} used the CXR images of COVID-19 infected and not infected patients to create a dataset to train a ResNet-50 deep learning model for COVID-19 automatic classification, and achieved 98\% accuracy. Zhang et al. \cite{Zhang2020Viral} used a deep learning model to distinguish between COVID-19 patients and pneumonia patients on the CXR image dataset of 70 patients diagnosed with COVID-19 and other pneumonia. The proposed deep learning model could reach a sensitivity of 90\% in the detection of COVID-19 pneumonia and a specificity of 87.84\% in the detection of non-COVID-19 pneumonia. Ozturk et al. \cite{Ozturk2020Automated} used a deep learning model, called DarkNet, to perform binary and multi-class classifications on CXR images of COVID-19 and other pneumonia patients. The whole dataset included 127 COVID19, 500 non-infected and 500 other infectious pathogens pneumonia cases. The results produced an accuracy score of 98\% for binary classification and 87.02\% for multi-class classification. Tang et al \cite{Tang2020Automated} and Shelke et al \cite{Shelke2020Chest} compared various classification architectures, including AlexNet \cite{Krizhevsky2012ImageNet}, VGG \cite{Simonyan2014Very}, ResNet \cite{He2015Deep}, Inception \cite{Szegedy2016Rethinking} and DenseNet \cite{Huang2018Densely}. The outstanding performance on the diagnostic accuracy of lung disease (Acc \textgreater 90\%) achieved in their studies on public CXR datasets showed that the deep learning-based classifier can accurately and effectively distinguish between different lung diseases, thereby providing potential benefits for improving patient care in clinical practice.

\subsection{Lung Segmentation}
\par Extracting different areas and lesions from CXR images can provide doctors with more relevant information to diagnose and quantify lung diseases \cite{Gordaliza2018Unsupervised}. The deep learning-based image segmentation methods have been used to detect the difference and the abnormality of areas in CXR images. In \cite{Gaal2020Attention}, Gaál et al. proposed an attention U-Net based adversarial architecture for lung segmentation using CXR images of COVID-19 infected and not infected patients. The method performed well on the CXR images of unseen datasets with different patient profiles, achieving an accuracy of 97.5\%. In \cite{Sarkar2020Detection}, COGNEXs Deep Learning Software-Vision Pro Deep Learning was used to classify and segment the regions of disease and lungs from the CXR dataset. The results were compared with various state-of-the-art Deep Learning models from the open-source community and achieved an F-score of 95.3\% . In \cite{Saeedizadeh2020COVID}, Saeedizadeh et al. proposed a deep learning framework for organ and abnormal area segmentation. The popular U-Net architecture through adding a connectivity promoting regularization term was used in the segmentation model. The trained model was able to achieve high accuracy results for detecting COVID-19 lung disease.

\subsection{Explainable Artificial Intelligence}
Due to the multiple nonlinear structures in deep learning networks, the deep learning-based classification methods are always considered as ''Blackbox'' approaches \cite{Buhrmester2019Analysis}. The state-of-the-art deep learning-based methods have been increasingly used in clinical prediction and healthcare, and have achieved reasonably high performance. However, these models do not provide the explanation to the classification results, leading to a limited understanding of the resulting prediction\cite{Zihni2020Opening}. So far there have been several visual explanation methods for the prediction results from deep learning-based models, such as saliency map \cite{adebayo2018sanity}, classification activation map (CAM) \cite{Zhou2016Learning}, Gradient-CAM(Grad-CAM)\cite{Selvaraju2016Grad-CAM:} and their variants. When using the saliency map for the prediction explanation, it is assumed that the positive gradient of a predicted category with respect to input image should ensure that category. The saliency map could provide a high-resolution gradient result with the same size as the image \cite{Simonyan2013Deep}. Class Activation mapping (CAM) was also a widely used explanation method for the object localisation extraction. In the CAM method, the top fully connected layers in a classification model are replaced with convolutional layers to keep the object positions, thus to discover the spatial distribution of discriminative regions for the predicted category. However, the CAM changes the model architecture which requires retraining the model, which limits the use of this method. Grad-CAM is a generalisation of the CAM method, it keeps the origin classification architecture and calculates the weights by pooling the gradient. This method has been widely used since it was proposed, because it can be used for all the deep learning-based classification models \cite{Selvaraju2016Grad-CAM:}. In \cite{Ahsan2020Study}, the author tested Local Interpretable Model-agnostic Explanations (LIME) and CAM methods to explain the state-of-the-art deep learning classification methods on CXR datasets. Brunese et al. \cite{Brunese2020Explainable} used the Grad-CAM method to automatically detect the areas of interest in the CXR images corresponding to the COVID-19 disease.

\section{The Proposed Method}
\par
\begin{figure}[h]
    \centering
    \includegraphics[width=0.45\textwidth]{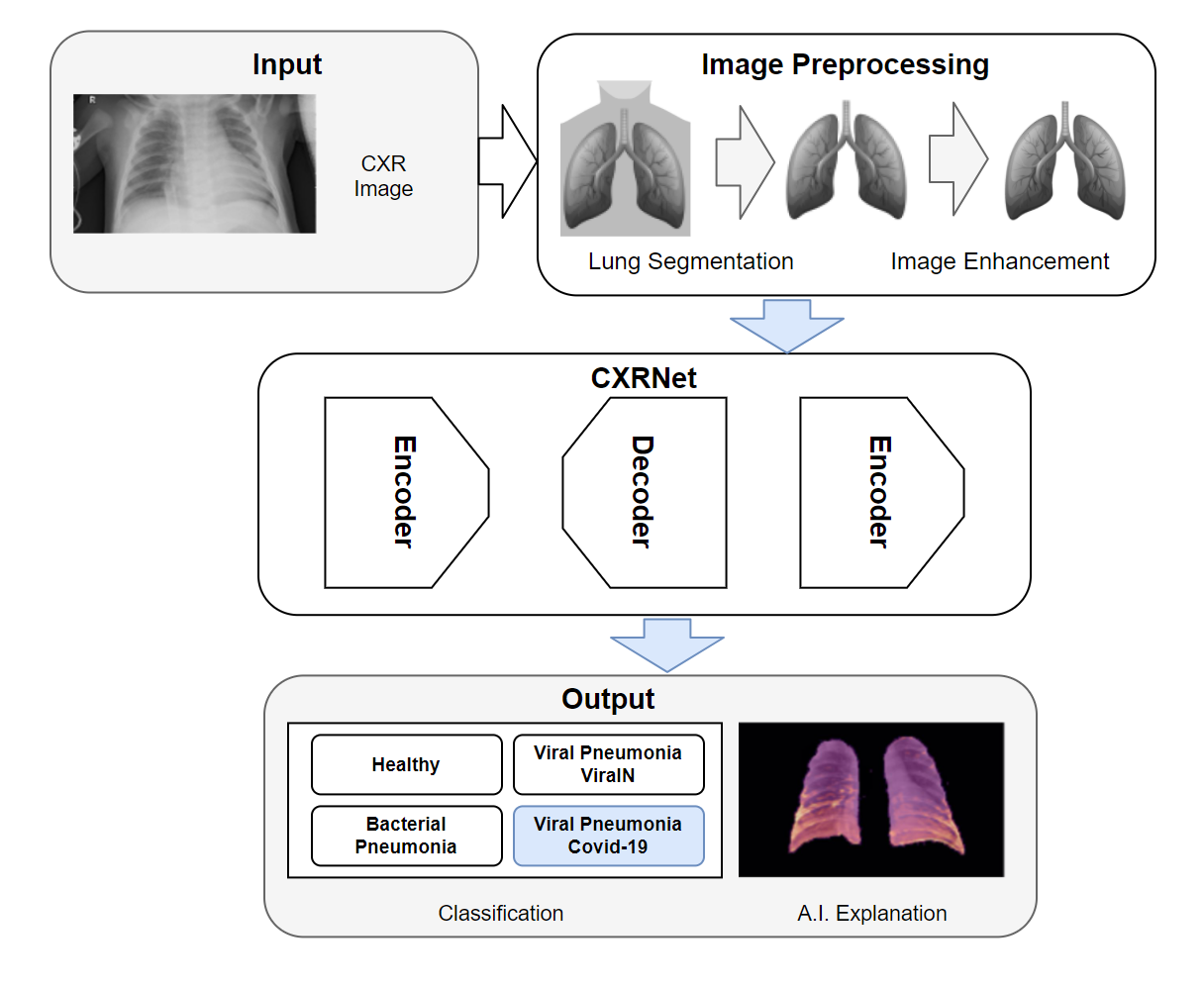}
    \caption{The overview of the framework}
    \label{FIG:1}
\end{figure}

 The aims of this study are (1) to classify a CXR image into one of the four classes: Health, Bacterial pneumonia, viral pneumonia of non infectious origin other than COVID-19 (ViralN) and COVID-19, and (2) to highlight the virus infected area on the image simultaneously. Therefore, we proposed a novel deep learning framework for classification and pixel-based visual explanation in an end-to-end manner. 
{Fig.~\ref{FIG:1}} provides the overview of the proposed solution consisting of two major components: 

\begin{enumerate}
  \item An Image Preprocessing module including Lung segmentation and Image enhancement steps. In the segmentation step, the lung areas are automatically extracted from the original image data based on the U-Net segmentation model. Then, the Contrast Limited Adaptive Histogram Equalization (CLAHE) is applied to the segmented images to enhance the contrast of grey CXR images.
  \item An Image classification and explanation model (CXRNet). The CXRNet is a convolutional neural network(CNN) based multi-task Encoder-Decoder-Encoder consisting of two jointly trained encoders for classification. The encoder is used for the classification task and the decoder is used to generate the visual explanation from the deep features extracted by the encoder. In this work, the first encoder is used to obtain the features of the original CXR images and the second one is used to extract the features of the representation image reconstructed from the decoder. The representation image between the two encoders acts as a proxy to visualize the most relevant areas associated with COVID-19 pneumonia diagnosis.
\end{enumerate}

\subsection{Image Preprocessing }
\label{sec:Imagepre}

\subsubsection{Lung Segmentation}
\begin{figure*}[h]
    \centering
    \includegraphics[width=0.95\textwidth]{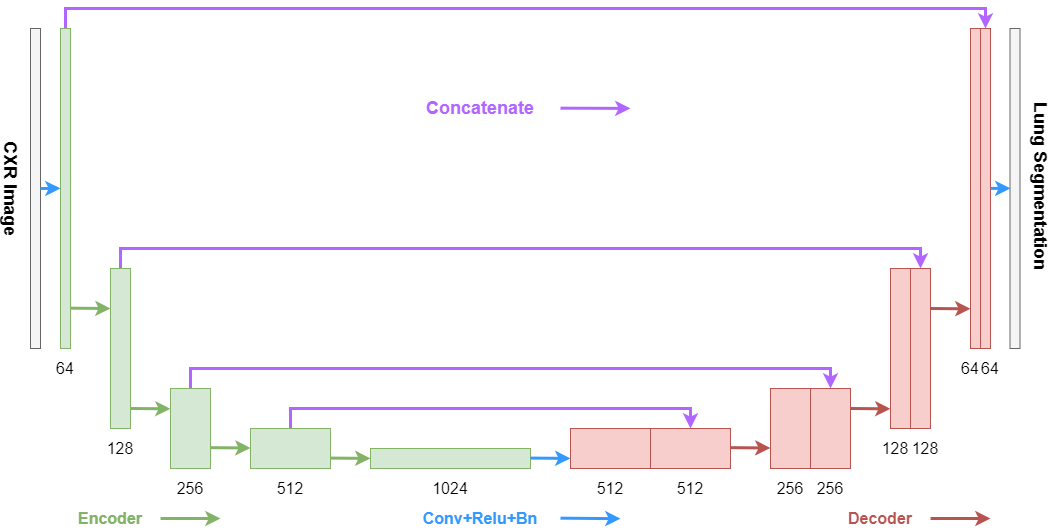}
    \caption{The U-Net segmentation model structure}
    \label{FIG:2}
\end{figure*}
\par Lung abnormality in CXR images is one important signal of the COVID-19 infection. To improve the classification performance, the lung area was extracted before performing image classification. In this study, a commonly used deep learning-based image segmentation model was selected to extract the lung area from the original CXR images. The U-Net, the structure of which is shown in {Fig.~\ref{FIG:2}}, was initially published for biomedical image segmentation \cite{Ronneberger2015U-Net:}. It concatenates the encoder feature maps to upsample feature maps from the decoder at each layer to form a ladder-like structure. Meanwhile, the architecture with skip concatenation connections allows the decoder to learn back relevant features that are lost during the pooling operations in the encoder. The architecture of U-net is simple and efficient, it consists of a contracting path to capture the context and a symmetric expanding path enabling precise localization. A hybrid loss mixed with Dice and Cross entropy (CE) loss was used as the loss function in the semantic segmentation networks in this study. The loss function in this work is defined as:
\begin{equation}
\operatorname{Loss}=\operatorname{Loss}_{C E}-\log \left(\text { Loss }_{\text {Dice }}\right)
\end{equation}

\begin{equation}
\operatorname{Loss}_{\text {Dice }}=\frac{2 \times T P}{2 \times T P+F P+F N}
\end{equation}
\par where TP, FP and FN are true positive, false positive, and false negative, respectively.

\subsubsection{Image Enhancement}
\par
\begin{figure}[h]
    \centering
    \includegraphics[width=0.45\textwidth]{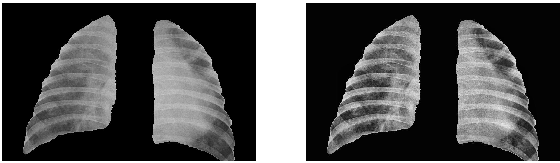}
    \caption{Example of CXR images before and after CLAHE}
    \label{FIG:3}
\end{figure}

 The grey X-rays images are typically low contrast, which makes their analysis challenging. Histogram equalization techniques can be used to enhance their contrast. The standard histogram equalization extends the most frequent intensity values to the higher range of the intensity domain [0,255], so that their cumulative distribution function (CDF) is closer to the uniform distribution. However, this method might over-amplify noise in near-constant regions. Instead, in this work, the Contrast Limited Adaptive Histogram Equalization (CLAHE) was chosen to enhance the contrast of CXR images. It clipped the histogram of an image at a predefined value before calculating the CDF, thus redistributing this part of the image equally among all the histogram bins. Applying the CLAHE to an X-ray image produces visually appealing results, as shown in {Fig.~\ref{FIG:3}}. 

\subsection{An Encoder-Decoder-Encoder Multitask Deep Neural Network for Disease Classification and Explanation (CXRNet)}
\begin{figure*}[h]
    \centering
    \includegraphics[width=0.8\textwidth]{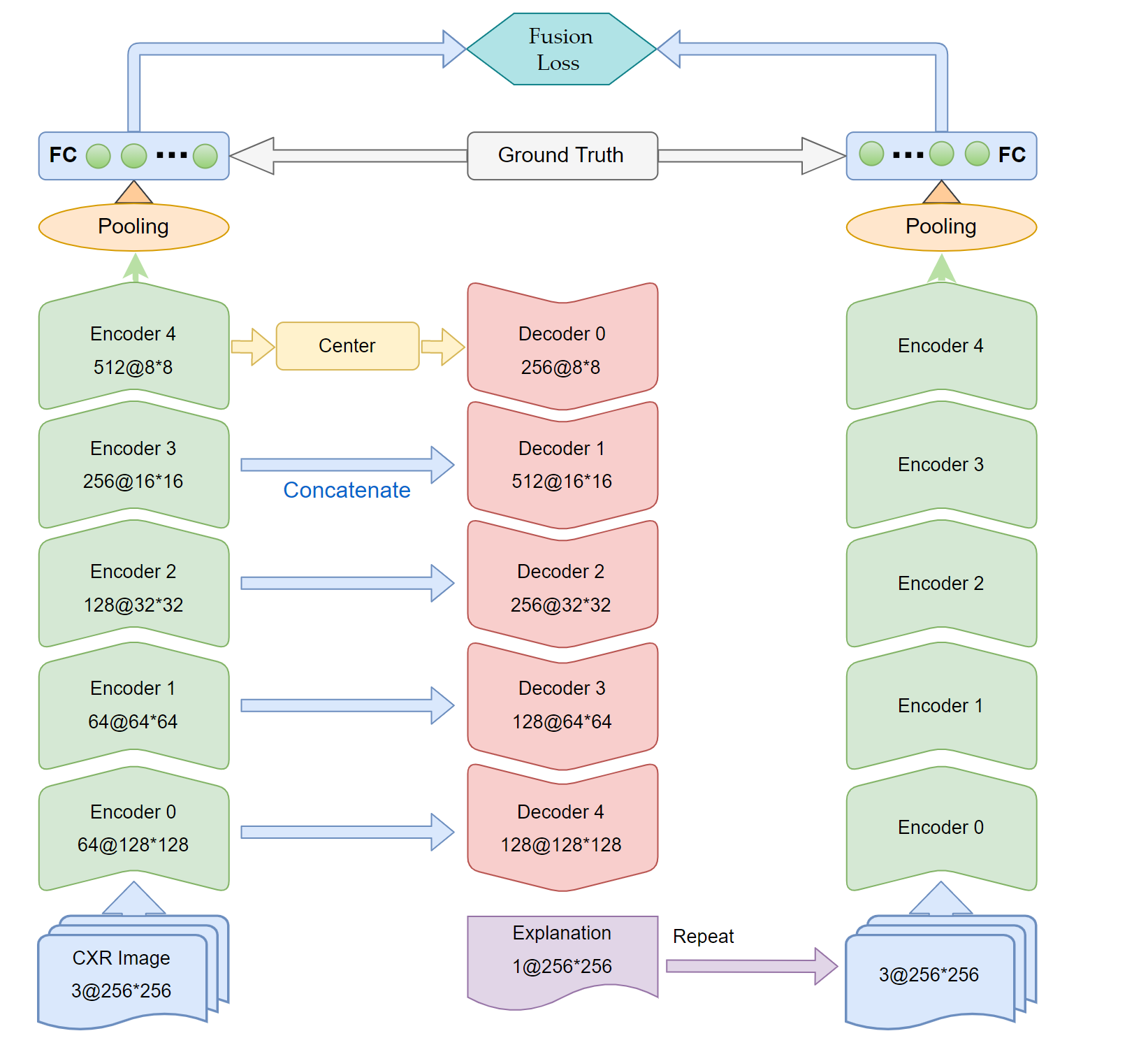}
    \caption{The architecture of the proposed classification and explanation model (CXRNet)}
    \label{FIG:4}
\end{figure*}

In this work, in order to generate a pixel-level visual explanation result alongside classification, we propose a multitask model of classification and visualization explanation, named CXRNet. The high-level conceptual framework of the proposal model is shown in {Fig.~\ref{FIG:4}}. It is a CNN-based Encoder-Decoder-Encoder structure, consisting of two encoders and one decoder.
\begin{enumerate}
  \item The first encoder in the model is a feature extraction network used to encode the input image into a certain feature representation, which is then fed into a classifier for identifying the disease types. The decoder is used to reconstruct an image with the same size as the input from the extracted feature representation.
  \item The second encoder is used to encode the reconstructed image from the decoder. It is jointly trained with the first encoder to make sure that the reconstructed image represents the most influential areas and features for classification. The second encoder shares the same weights as the first encoder, and uses a fusion loss combining the loss functions of both encoders. During the model training, the parameters of the decoder are iteratively updated to make the generated image keep the most relevant features for identifying the disease.
\end{enumerate}

\paragraph{An Encoder-Decoder Structure}

\par In this work, we designed an encoder-decoder-encoder network architecture to perform multitask learning for both disease classification and visual explanation tasks. A U-NET like architecture with skip concatenation connections is adopted, as shown in {Fig.~\ref{FIG:4}}. The network architecture consists of 5 encoder layers and 5 decoder layers. The skip concatenation connections are added between each encoder and decoder layer to allow the decoder at each stage to learn back relevant features that are lost when performing pooling operations in the encoder. The decoder adopts a structure similar to the encoder by adding an up-sampling layer to extend the deep feature to the original size. The aim of the up sampling (U) is to add pixels around the existing pixels and also in-between to eventually reach the desired size. For each decoder block, the upscale is set to 2 to make sure that the output has the same size as the forward encoder output. In this model we use an improved method, known as pixel shuffle with ICNR initialization, which makes the gap filling between the pixels much more effectively and avoids generating checkerboard artifacts \cite{Sugawara2018Super-Resolution}. 
\par
\begin{figure}[h]
    \centering
    \includegraphics[width=0.45\textwidth]{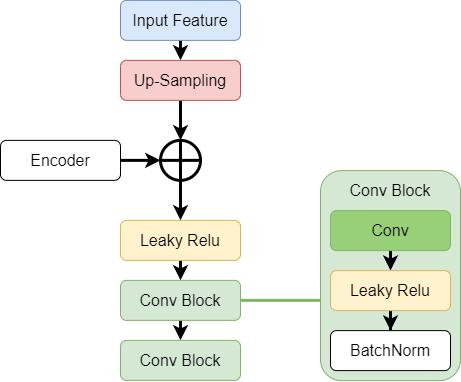}
    \caption{The Decoder structure}
    \label{FIG:5}
\end{figure}

 The structure of the decoder is shown in {Fig.~\ref{FIG:5}} and can be formulated as follows:

\begin{equation}
\text { Decoder }_{n}=F\left(U\left(\text { Decoder }_{n-1}\right)+\text { Encoder }_{4-n}\right)
\end{equation}

\par where U(Decoder) denotes an up-sampling operation on the decoder outputs, and F is the combination operation on the concatenation of the feature outputs from the up-sampling operation and the encoder. The F operation contains a rectified linear unit (ReLU) and two convolution blocks. Each convolution block consists of 3x3 D convolution layers, batch normalization and ReLU. In this work, the leaky ReLU activation function \cite{maas2013rectifier} with the negative slope of 0.1 is used to replace the standard ReLU activation in the decoder part. This allows a small gradient when the unit is not active, which has proven to outperform the original ReLU, especially in an encoder-decoder architecture\cite{Antic2019Deep,Lehtinen2018Noise2Noise:,Xu2015Empirical}. The equation of leaky ReLU is defined as follows:
\begin{equation}
\text { LeakyReLU }(x)=\left\{\begin{aligned}
x \times \text { slope }, & x<0 \\
x, & x \geq 0
\end{aligned}\right.
\end{equation}

\paragraph{Encoders for Classification}

\par
\begin{figure*}[h]
    \centering
    \includegraphics[width=0.9\textwidth]{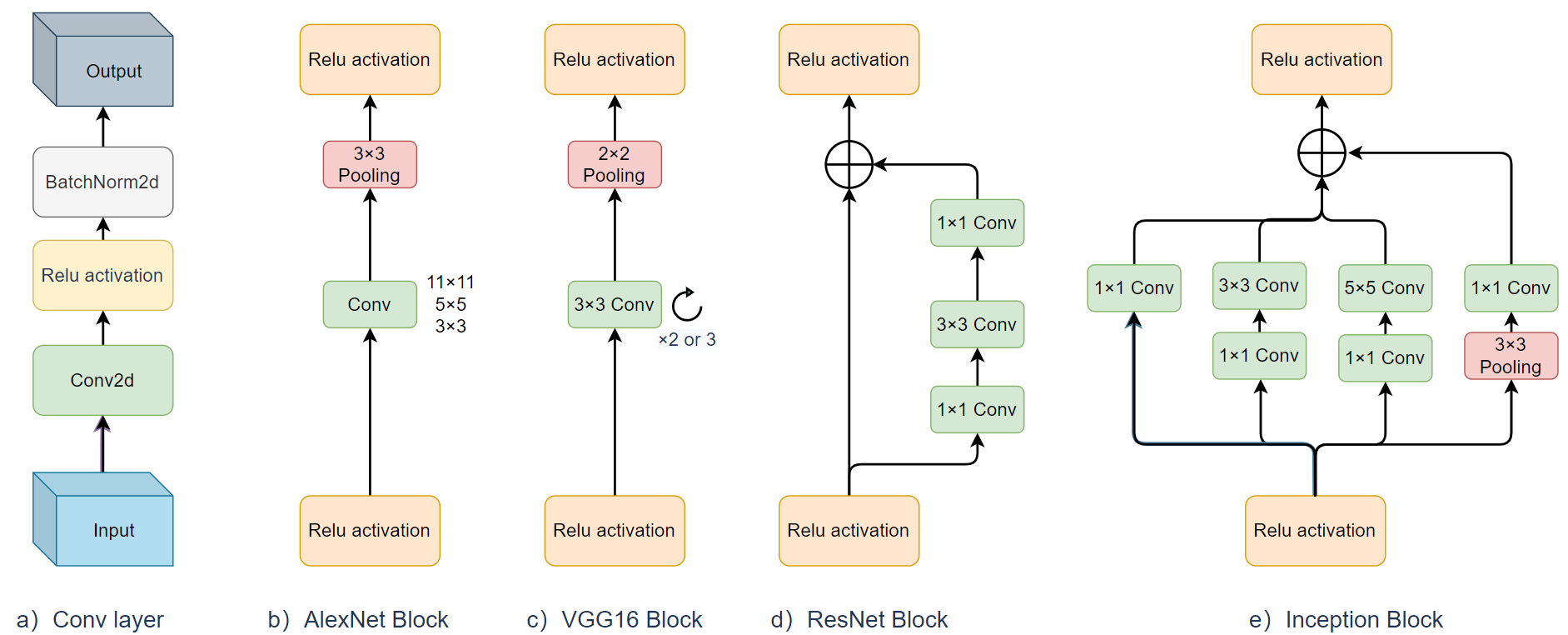}
    \caption{The schematic diagrams of AlexNet, VGG16, ResNet and Inception encoder block}
    \label{FIG:6}
\end{figure*}

 The proposed model has two encoders. The first one directly extracts features from the input image, and the second is to extract the features from the reconstructed image. The two encoders share the same weights. After the feature extraction by the two encoders, the extracted feature maps are transformed into the input of a classifier for disease type identification through an average pooling layer and a fully connected layer. The classifier can adopt most of existing CNN classification architectures. In this work, we evaluated four most commonly used encoder architectures, AlexNet \cite{Krizhevsky2012ImageNet}, VGG16 and 19 \cite{Simonyan2014Very}, ResNet 34 and 50 \cite{He2015Deep}, Inception-V3 \cite{Szegedy2016Rethinking}. Their structures are shown in {Fig.~\ref{FIG:6}}.

\paragraph{Fusion Loss}
\par The loss function of the network is designed to minimize jointly the losses of the two classifiers (i.e. encoder). The cross-entropy loss is selected as the classification loss function. The fusion loss function in this work is defined as:

\begin{equation}
\operatorname{Loss}=\operatorname{Loss}_{C E 1}+\operatorname{Loss}_{C E 2}
\end{equation}

\begin{equation}
\operatorname{Loss}_{C E}=-\sum_{i}^{C} t_{i} \log \left(p_{i}\right)
\end{equation}

\par The ${Loss}_{CE1}$ and ${Loss}_{CE2}$ are the loss value of the first encoder and the second encoder, respectively, and $\ p_i$ is the probability of class $i$.

\section{Experimental evaluation}

\subsection{Dataset description}

In this work, 6499 CXR images were used for model training. Among them, there were 636 cases of COVID-19. The dataset was collected from multiple public and private sources: 5,863 cases from \cite{kermany2018labeled}, 116 COVID-19 cases from \cite{Maguolo2020Critic,Tartaglione2020Unveiling}, 479 COVID cases from \cite{cohen2020covid,cohen2020covidProspective} and 41 images from COVID-19 patients on the Critical Care unit at Whiston Hospital, St Helen's \& Knowsley NHS Trust, UK. 

\begin{figure}[h]
    \centering
    \includegraphics[width=0.45\textwidth]{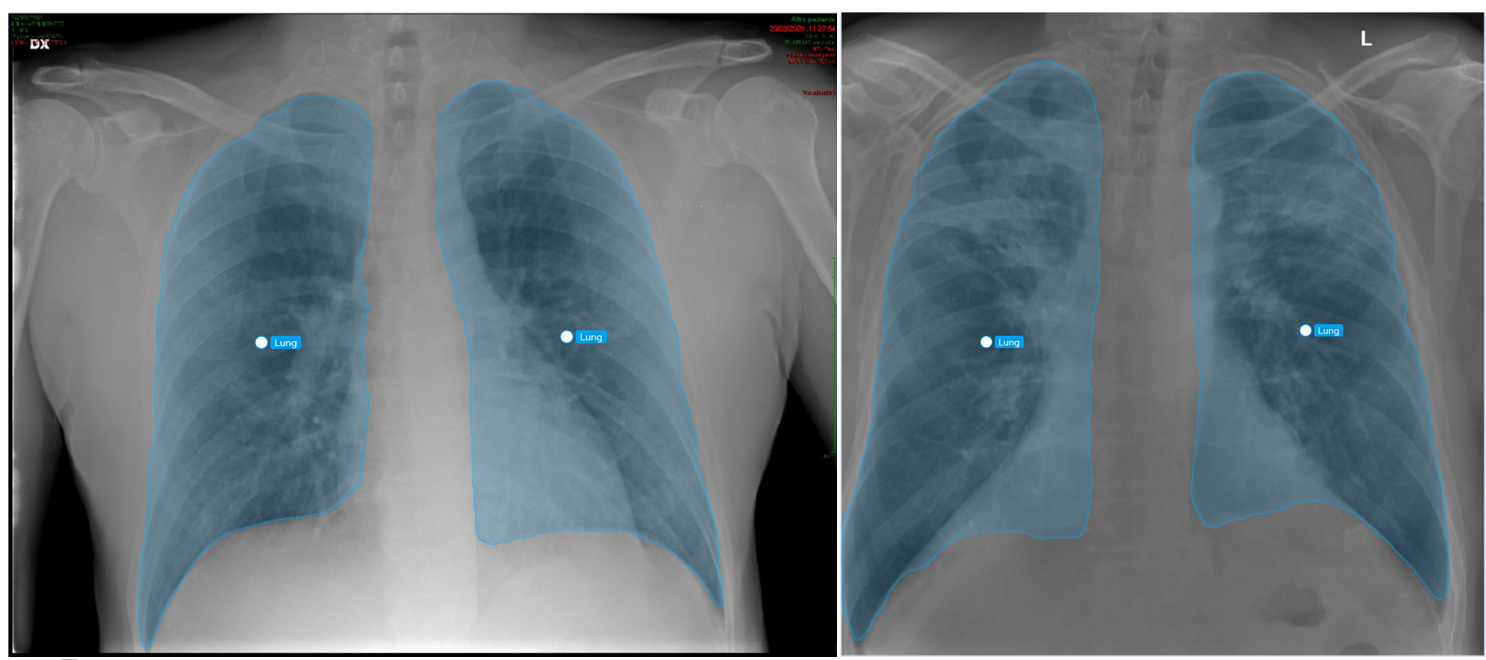}
    \caption{Two examples of pixel-level polygonal lung segmentation on the CXR images. The left one is the segmented result of a CXR image from a patient without Covid-19, and the right one is the segmented result of a CXR image from a Covid-19 patient. The lung segmentation results are superimposed on the raw CXR images.}
    \label{FIG:7}
\end{figure}

All the images were categorized into three classes: Health, Bacterial Pneumonia and Viral Pneumonia. The COVID-19 CXR images were labelled as Viral Pneumonia. In this work we split the cases of Viral Pneumonia into two groups, non-COVID-19 viral pneumonia (ViralN) and COVID-19. The images from V7lab provided a pixel-level polygonal lung segmentation on the CXR images \cite{2020v7labs/covid-19-xray-dataset}. {Fig.~\ref{FIG:7}} shows some examples of the lung segmentation labels and the image category. 

\subsection{Model Training and Performance Evaluation Metrics}

In this work, the model was implemented based on the Pytorch 1.6 and executed on a PC with an Intel (R) Xeon (R) CPU E5-2650, NVIDIA TITAN X (Pascal) and 64GB memory. We randomly selected 80\% of samples for training and used the rest to evaluate the model performance. Adam (A Method for Stochastic Optimisation) with a batch size of 12 was chosen as the optimisation algorithm in the model. We initially set a base learning rate of $1 \times 10^{-4}$. The base learning rate was decreased to $1 \times 10^{-6}$ with the increased iterations.
To evaluate the performance of our segmentation model, Intersection over Union (IoU) a commonly used metric, was used. It is defined as:

\begin{equation}
I o U=\frac{\text { Area of Overlap }}{\text { Area of Union }}
\end{equation}
\par That is, the area of overlap between the predicted segmentation and the ground truth divided by the area of union between the predicted segmentation and the ground truth. IoU ranges in value from 0 to 1. 
\par To evaluate the classification performance of the proposed architecture, Recall, Precision, ${F1}_{score}$ , Average Accuracy and Confusion Matrix are selected as the accuracy performance metrics. The average accuracy can be calculated from the true positive (TP), the true negative (TN), the false positive (FP) and the false negative (FN). The metrics are calculated as follows:
\begin{equation}
\text { Recall }=\frac{T P}{T P+F N}
\end{equation}

\begin{equation}
\text { Precision }=\frac{T P}{T P+F P}
\end{equation}

\begin{equation}
F 1_{\text {score }}=\frac{\text { Recall } \times \text { Precision }}{\text { Recall }+\text { Precision }} \times 2
\end{equation}

\begin{equation}
\text { Accuracy }=\frac{T P+T N}{T P+T N+F P+F N}
\end{equation}

\par Furthermore, Probability Perturbation Curve (PPC) is calculated to evaluate the quality of visual explanation. The essence of PPC is to proportionally remove the ’important’ regions in an image to produce perturbations of variables to the original image and then evaluate how the classification performance responds to the changes of the different portions of removals (perturbations). For instance, when removing the large portion of important regions from an image, the classification performance is supposed to be lower than without removal. Therefore, the greater the perturbation to the classifier, the more important the removed region. Here, the value of the pixels in the visualisation result determines their importance for the classification. We started from removing the pixels with high values in the explanation result to generate the perturbed images. Then, each perturbed image was fed to the trained classifier to get the probability of the corresponding type. The curve was plotted following the change of points (x, p(f(x))) during the pixel removal procedure. x started from 1 to 0, and f(x) denoted the removal procedure in which all the pixels with a value of x were replaced with zeros. So, f(0) means no change to the original image, and f(1) means all the pixels with a value of 1 (the most important regions detected by models) will be replaced with zeros. $p(f\left(x\right))$ is the probability calculated from the classifier based on the processed image. This approach was used for all COVID-19 images and the probabilities were averaged to produce the curve. The area over probability perturbation curve (AOPPC) was used to evaluate the quality of visual explanation and can be formulated as follows: 
 %  In this work, we removed the 'important' regions of image to produce perturbations of variables to the original image. Naturally, a classifier will be more confused by a larger perturbation. Therefore, the greater the perturbation to the classifier, the more important the removed region. Here, the value of the pixels in the visualisation result determines their importance for the classification. We started from removing the pixels with high values in the explanation result to generate the perturbed images. Then, each perturbed image was fed to the trained classifier to get the probability of the corresponding type. The curve was plotted following the change of points (x, p(f(x))) during the pixel removal procedure. x started from 1 to 0, and f(x) denoted the removal procedure in which all the pixels with a value of x were replaced with zeros. So, f(0) means no change to the original image, and f(1) means all the pixels with a value of 1 will be replaced with zeros. $p(f\left(x\right))$ is the probability calculated from the classifier based on the processed image. This approach was used for all COVID-19 images and the probability were averaged to produce the curve. The area under curve (AUC) was used to evaluate the quality of visualization explanation and can be formulated as follows: 
  
\begin{equation}
AOPPC=\frac{1}{10} \sum_{i=1}^{n} \sum_{x=0.1}^{1}(p(f(0))-p(f(x)))
\end{equation}

\subsection{Experimental Evaluation}

\subsubsection{Lung segmentation}

\par
\begin{figure}[h]
    \centering
    \includegraphics[width=0.45\textwidth]{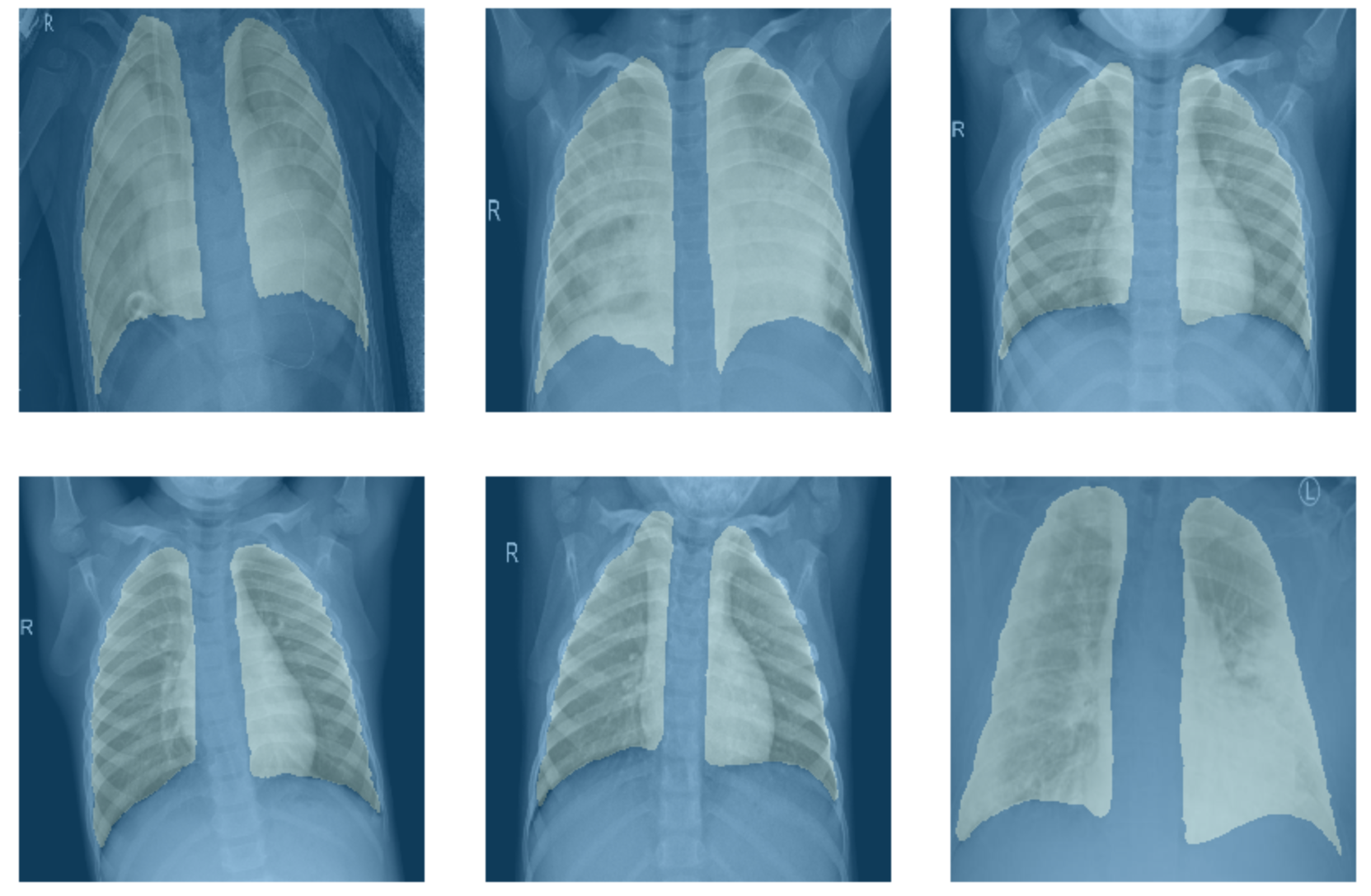}
    \caption{Six examples of Lung area segmentation results}
    \label{FIG:8}
\end{figure}

 The lung areas in CXR images are the regions of interest (ROI) for the COVID-19 diagnosis. In this paper, we firstly trained a U-Net semantic segmentation model to extract the lung areas from the original CXR image. The IoU accuracy reached 0.925, close to the accuracy from other works\cite{Gaal2020Attention,Saeedizadeh2020COVID,Sarkar2020Detection}. {Fig.~\ref{FIG:8}} shows six examples of lung area segmentation results.

\subsubsection{Lung disease diagnosis}

\par
\begin{figure}[h]
    \centering
    \includegraphics[width=0.45\textwidth]{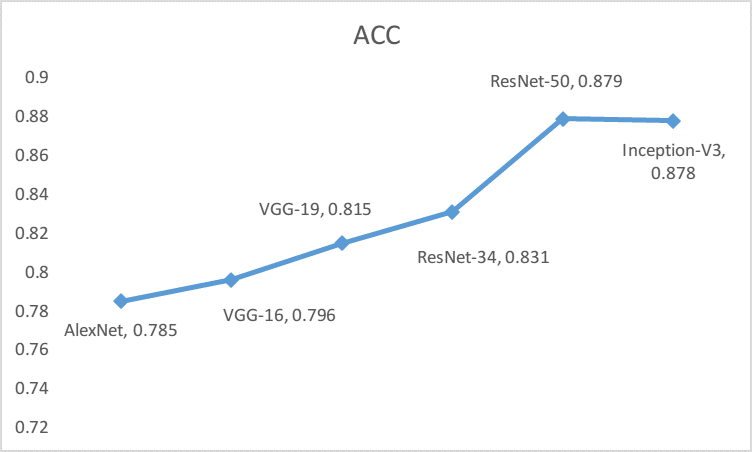}
    \caption{Classification Accuracy Comparison of six different CNN architectures for the encoder}
    \label{FIG:9}
\end{figure}

 Firstly, we evaluated the classification performance of the proposed model on original CXR images. Six most commonly used CNN architectures were considered for the encoder. They included AlexNet, VGG-16, VGG-19, ResNet-34, ResNet-50 and Inception-V3. {Fig.~\ref{FIG:9}} shows a comparison of the average accuracy (ACC) when different CNN architectures are employed for the encoder. Among them, the models in which the encoder used ResnNet-50 or Inception-V3 architectures achieved the best average accuracy. A better performance of the model with ResNet-50 (50 layers) over the model with ResNet-34 (34 layers) shows that a deeper architecture may boost the performance. Due to the limitation of GPU memory, we did not choose deeper architectures than ResNet-50.  

\begin{figure}[h]
    \centering
    \includegraphics[width=0.45\textwidth]{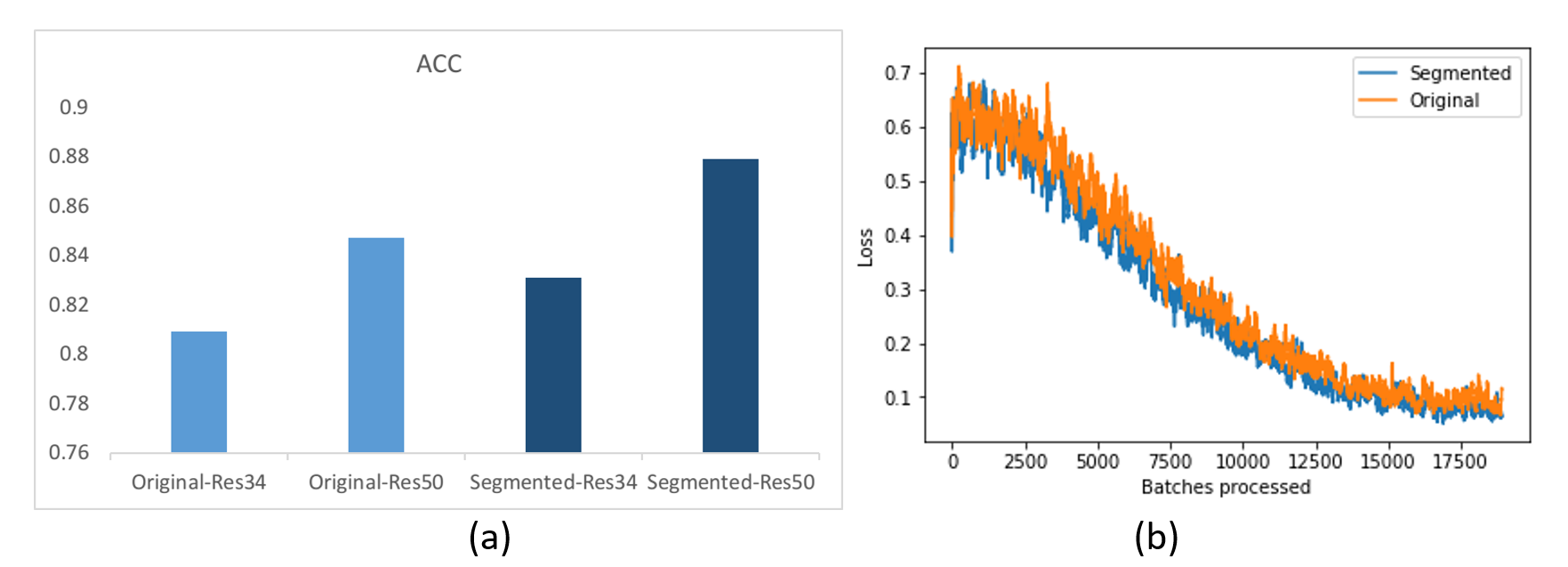}
    \caption{a) A comparison of the classification accuracy using original image or lung segmented image; b) A comparison of the training loss trend of Resnet50 when using original and segmented images}
    \label{FIG:10}
\end{figure}

\begin{figure}[h]
    \centering
    \includegraphics[width=0.45\textwidth]{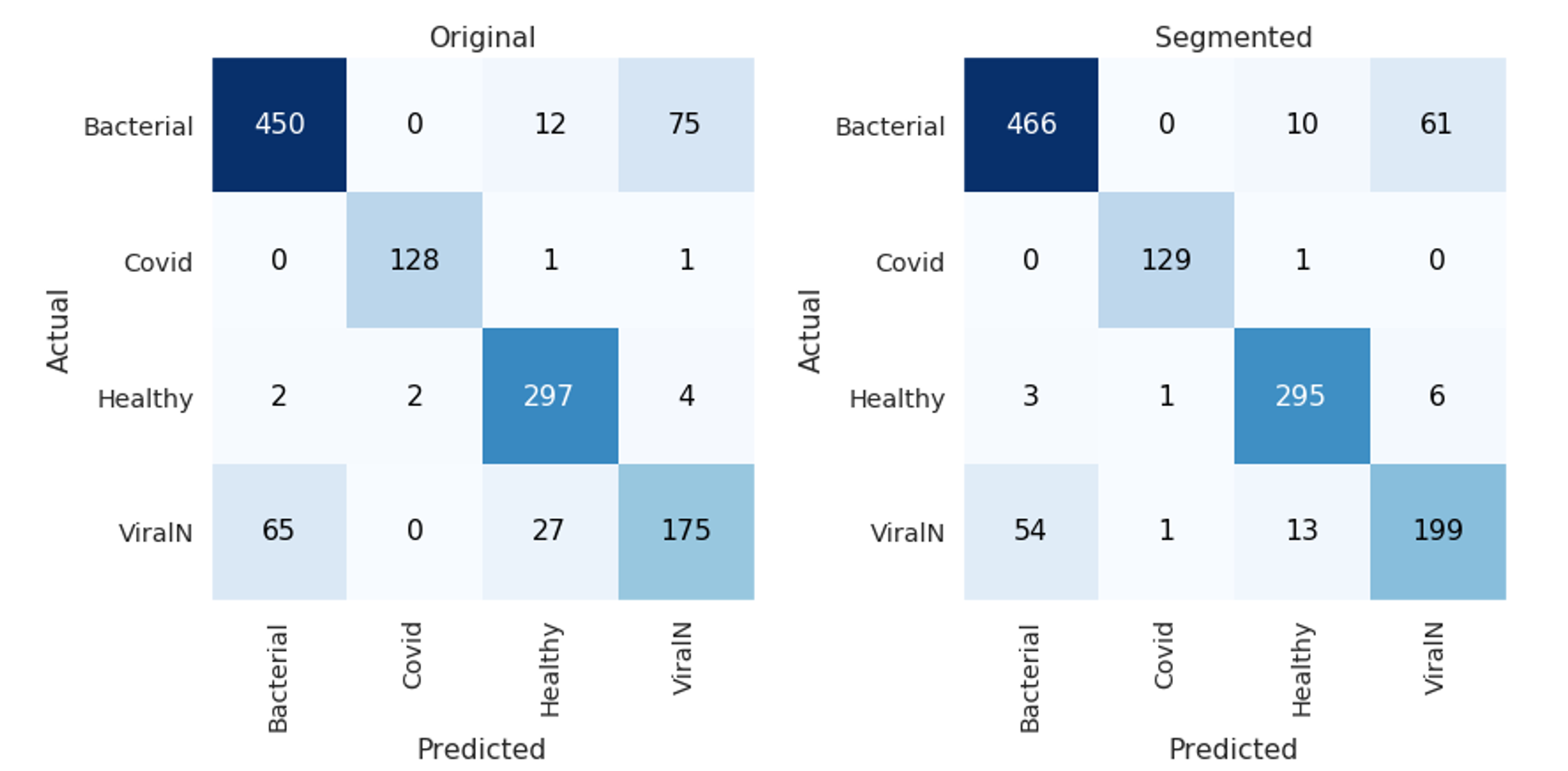}
    \caption{Confusion matrix by the ResNet50-based model trained on Original and Lung Segmented CXR images}
    \label{FIG:11}
\end{figure}

Secondly, we investigated the effect of lung segmentation on the classification performance. {Fig.~\ref{FIG:10}} a) shows a comparison of the classification accuracy when the models are trained on original images and lung segmented images, respectively. Two different depth ResNet models (34 and 50 layers) are compared as well. The results show that the performance of the model has been improved when trained on lung segmented images instead on original images. Increasing the depth of the model can also improve the classification performance. {Fig.~\ref{FIG:10}} b) shows the training loss curves of ResNet50 based encoder models when trained on original and segmented images. It can be found that the loss of the model trained on segmented images is lower than that of the model trained on original images. This indicates that the segmented images may be easier for the model to predict than the original images.
The performance of models respectively trained on original images and lung segmented images can be further evaluated with the confusion matrix, as shown in {Fig.~\ref{FIG:11}}. Each entry in a confusion matrix denotes the number of predictions made by the model where it classifies the classes correctly or incorrectly. Each column shows the predicted results of the model for each class. The Precision, Recall and F1 score are provided in Table~\ref{table:1} for the ResNet50-based encoder using original and segmented CXR images. From the Confusion Matrix, we can find that the model performs well on the detection of COVID-19. 128 out of 130 and 129 out of 130 cases are correctly predicted by the model trained on original and lung segmented CXR images, respectively. When the model was trained on the segmented images, it achieved an average accuracy of 0.879 for all classes and an F1-score of 0.989 for COVID-19. However, bacterial and normal Viral pneumonia were harder to be distinguished by the model. 61 out of 537 bacterial pneumonia CXR images were incorrectly predicted as non-COVID-19 viral pneumonia, while 54 out of 267 non-COVID-19 pneumonia were incorrectly predicted as bacterial pneumonia.

\begin{table}[h]
\caption{Results of classification for AD vs. NC and pMCI vs. sMCI}\label{table:1}
\centering
\resizebox{0.45\textwidth}{!}{%
\begin{tabular}{rrrrrrr}
\rowcolor[HTML]{FFFFFF} 
\multicolumn{1}{l}{\cellcolor[HTML]{FFFFFF}} & \multicolumn{3}{c}{\cellcolor[HTML]{FFFFFF}Original}                                                                                                              & \multicolumn{3}{l}{\cellcolor[HTML]{FFFFFF}Segmented}                                                                                                             \\ \hline
\rowcolor[HTML]{F2F2F2} 
\multicolumn{1}{l}{\cellcolor[HTML]{FFFFFF}} & \multicolumn{1}{l}{\cellcolor[HTML]{F2F2F2}Precision} & \multicolumn{1}{l}{\cellcolor[HTML]{F2F2F2}Recall} & \multicolumn{1}{l}{\cellcolor[HTML]{F2F2F2}F1-score} & \multicolumn{1}{l}{\cellcolor[HTML]{F2F2F2}Precision} & \multicolumn{1}{l}{\cellcolor[HTML]{F2F2F2}Recall} & \multicolumn{1}{l}{\cellcolor[HTML]{F2F2F2}F1-score} \\
\cellcolor[HTML]{FFFFFF}Healthy               & 0.881                                                 & 0.974                                              & 0.925                                                & 0.925                                                 & 0.967                                              & 0.946                                                \\
\rowcolor[HTML]{F2F2F2} 
\cellcolor[HTML]{FFFFFF}Bacterial            & 0.87                                                  & 0.838                                              & 0.854                                                & 0.891                                                 & 0.868                                              & 0.879                                                \\
\cellcolor[HTML]{FFFFFF}ViralN               & 0.686                                                 & 0.655                                              & 0.67                                                 & 0.748                                                 & 0.745                                              & 0.747                                                \\
\rowcolor[HTML]{F2F2F2} 
\cellcolor[HTML]{FFFFFF}COVID-19                & 0.985                                                 & 0.985                                              & 0.985                                                & 0.985                                                 & 0.992                                              & 0.989                                                \\ \hline
\cellcolor[HTML]{FFFFFF}Accuracy             & \multicolumn{3}{c}{0.847}                                                                                                                                         & \multicolumn{3}{c}{0.879}                                                                                                                                         \\ \hline
\end{tabular}
}
\end{table}

\subsubsection {Model Explainability}
 
The heatmap of classification activation and the AOPPC were used to evaluate the performance of our proposed model on visual explanation and classification. It was also compared with two most commonly used DCNN explanation methods: Saliency map and Grad-CAM.

\begin{figure}[h]
    \centering
    \includegraphics[width=0.45\textwidth]{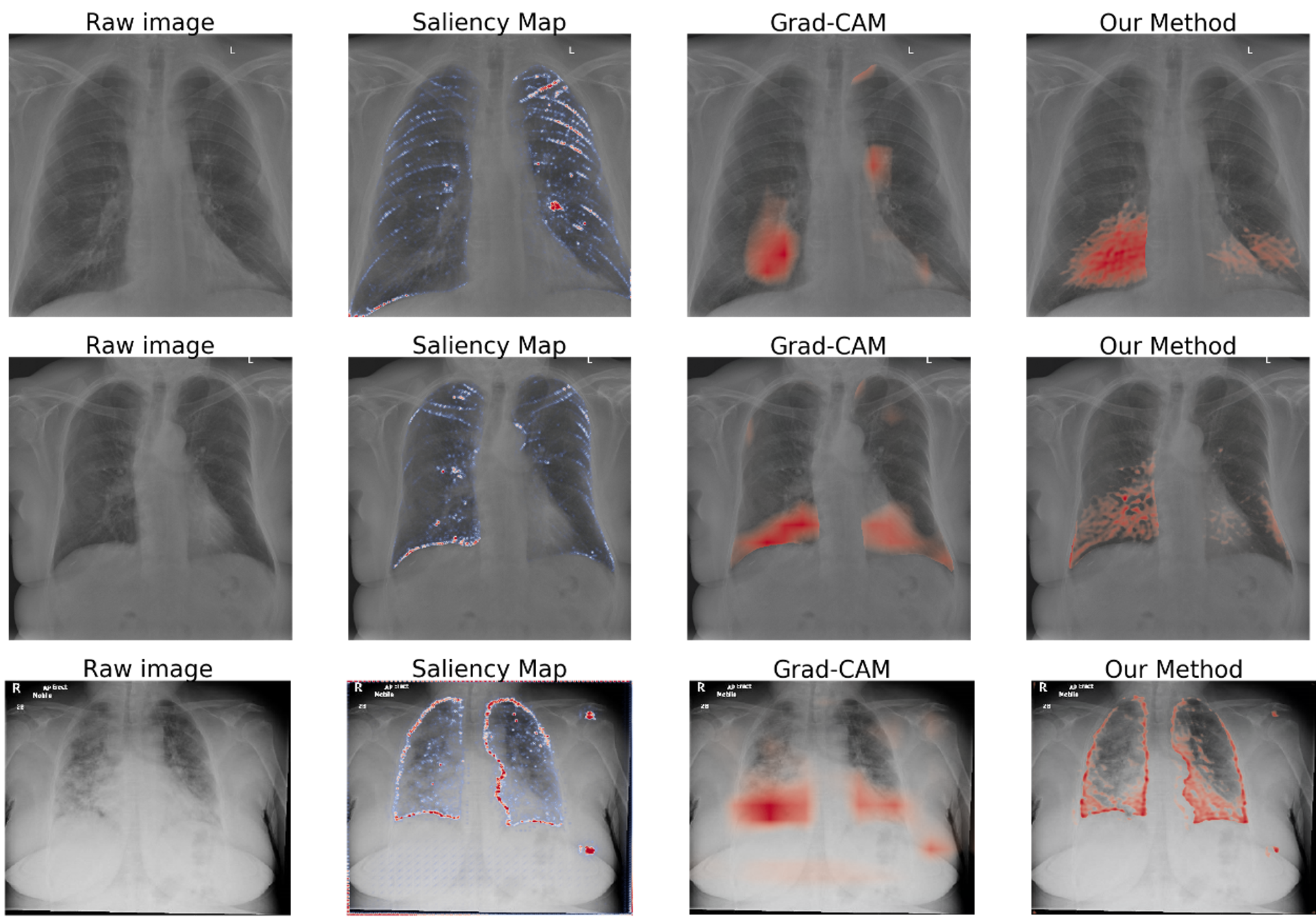}
    \caption{Three typical examples of class activation heatmap from three visualisation algorithms (a) Saliency Map (b) Grad-CAM (c)Our method}
    \label{FIG:12}
\end{figure}

\par {Fig.~\ref{FIG:12}} presents three typical examples of class activation heatmaps from the existing methods (Saliency map and Grad-CAM) and our proposed approach. Compared to the Saliency map and Grad-CAM approaches, our proposed algorithm provides shaper heatmaps due to its capability to generate the pixel-level results. The saliency map results are more noisy because it uses the gradients to measure the sensitivities of pixels rather than their contributions to classification. Although the Grad-CAM can localise globally the important regions, it has missed some areas identified by the proposed method. The missed identification may be due to the low-resolution features used by the Grad-CAM when propagating the contributions of class activation from the last convolution layer in the models.

\begin{figure}[h]
    \centering
    \includegraphics[width=0.45\textwidth]{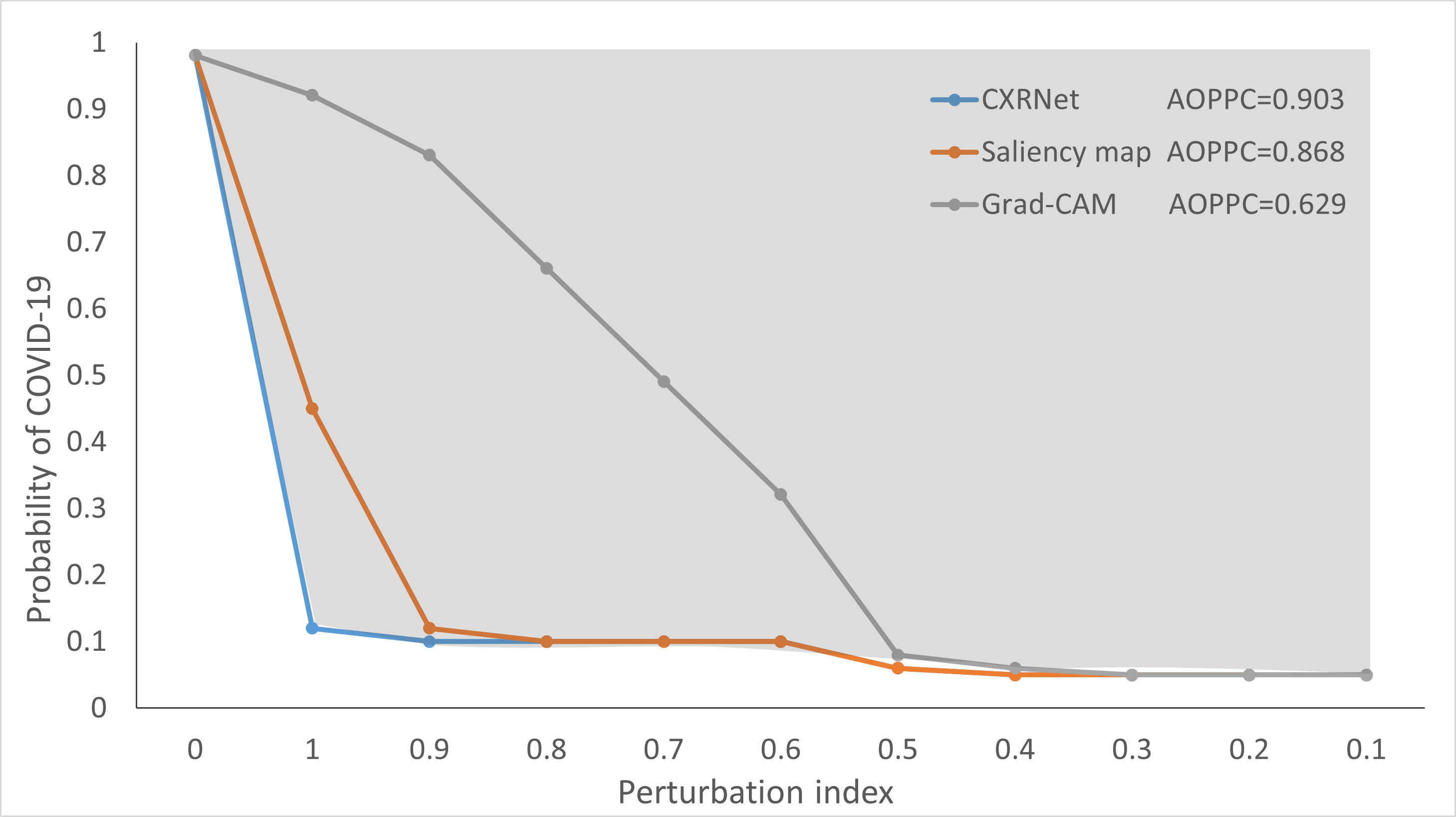}
    \caption{The probability perturbation curves (PPC) and area over probability perturbation curve (AOPPC)  of three explanation methods. The perturbation index in x axis shows the level of perturbation on images by removing the regions with a pixel value of x. The y axis presents the probability of COVID-19 predicted by the trained classifier on the perturbed images.}
    \label{FIG:13}
\end{figure}

\begin{figure}[h]
    \centering
    \includegraphics[width=0.45\textwidth]{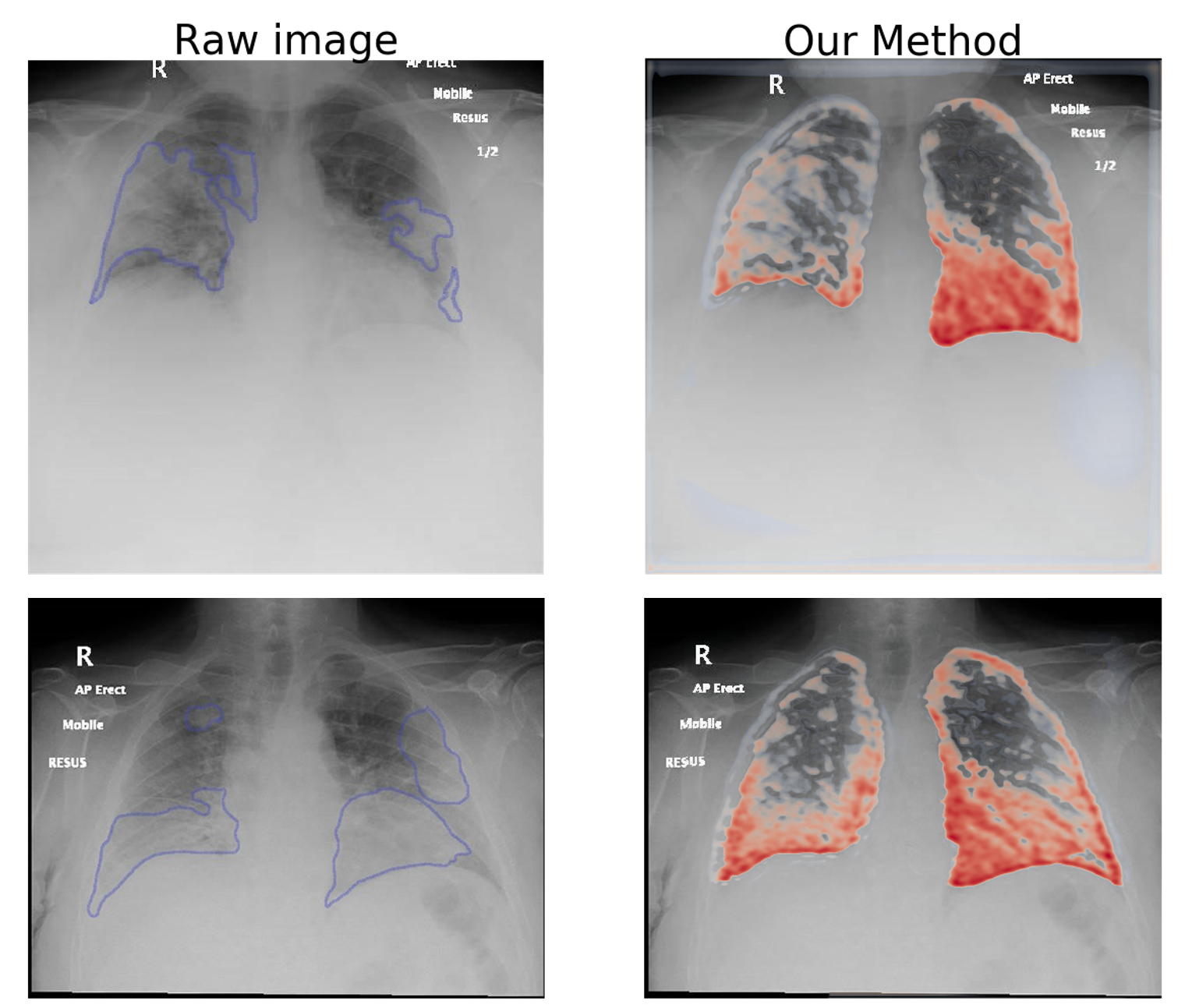}
    \caption{The CXR image with COVID-19 pneumonia. On the left column, the blue polygons on the images marked the ground glass opacities, the bilateral dense peripheral consolidation and the linear opacity area annotated by radiologists. The right column presented the heatmap from our method. The ground glass opacity, linear opacity and consolidation areas were all highlighted by our model.}
    \label{FIG:14}
\end{figure}

{Fig.~\ref{FIG:13}} shows the PPCs from three visualization approaches. The AOPPCs from the Salient Map, Grad-Cam and our methods are 0.629, 0.868 and 0.903, respectively. Our method has a higher classification score for COVID-19 detection.
The covid-19 pneumonia can increase the density of lungs, which can be seen as whiteness on Chest radiography. Therefore, multi-focal ground-glass opacity, linear opacities, and consolidation can be seen as evidence for the existence of COVID-19 infection \cite{Cleverley2020role}. {Fig.~\ref{FIG:14}} shows two examples of our heatmap results along with manual annotations on CXR images of COVID-19 pneumonia patients. On the left column, the blue polygons on the raw images mark the glass opacities, the bilateral dense peripheral consolidation and the linear opacity area annotated by two radiologists. The right column shows the heatmaps by our method. As shown in {Fig.~\ref{FIG:14}}, all the ground-glass opacity and linear opacity areas annotated by the radiologists are highlighted by our model.

\section{Conclusion}

In this work, we proposed a pixel level explainable classification model (CXRNet) for COVID-19 pneumonia diagnosis. The architecture is based on an encoder-decoder-encoder architecture, which enables the multitask learning for accurate and explainable disease classification. The experiments demonstrated that the proposed method achieved a reasonably high accuracy on classification using lung segmented CXR images. The model achieved an average accuracy of 0.879, and the Precision, Recall and F1-score of COVID-19 were 0.985, 0.992 and 0.989, respectively. Moreover, the model also generated a sharper and more precise visualisation images, compared to the Salient Map and Grad-Cam visualisation approaches. The explanation results can also highlight the discriminant regions, which helps to explain the classifier's decision and diagnosis of COVID-19 pneumonia. Meanwhile, we trained a segmentation model for lung area segmentation with a satisfactory accuracy of 0.925 which has shown that using lung segmented images can improve the classification accuracy.

%\section*{Acknowledgment}

%The work reported in this paper has formed part of the project by 
% Royal Society - Academy of Medical Sciences Newton Advanced Fellowship (NAF$\backslash$R1$\backslash$180371).

\ifCLASSOPTIONcaptionsoff
  \newpage
\fi



\bibliographystyle{IEEEtran}
\bibliography{covid}
\end{document}